\documentclass[jkps,preprint,fleqn,showpacs,showkeys]{revtex4}

\usepackage{graphicx}
\usepackage{amssymb}
\usepackage{amsmath}
\usepackage{bm}
\usepackage{multirow}
\usepackage{setspace}
\usepackage{graphicx}

\begin{document}

\setcounter{page}{0}

\title[]{$^{24}$Mg($p$, $\alpha$)$^{21}$Na reaction study for spectroscopy of $^{21}$Na}

\author{S.~M.~\surname{Cha}}
\affiliation{Department of Physics, Sungkyunkwan University, Suwon 440-746, Korea,}
\author{K.~Y.~\surname{Chae}}
\email{kchae@skku.edu}
\thanks{Fax: +82-31-290-7055}
\affiliation{Department of Physics, Sungkyunkwan University, Suwon 440-746, Korea,}
\author{S.~\surname{Ahn}}
\affiliation{Department of Physics and Astronomy, University of Tennessee, Knoxville, Tennessee 37996, USA,}
\author{D.~W.~\surname{Bardayan}}
\affiliation{Department of Physics, Notre Dame University, Notre Dame, Indiana 46556, USA,}
\author{K.~A.~\surname{Chipps}}
\affiliation{Department of Physics, Colorado School of Mines, Golden, Colorado 80401, USA,}
\author{J.~A.~\surname{Cizewski}}
\affiliation{Department of Physics and Astronomy, Rutgers University, New Brunswick, New Jersey 08903, USA,}
\author{M.~E.~\surname{Howard}}
\affiliation{Department of Physics and Astronomy, Rutgers University, New Brunswick, New Jersey 08903, USA,}
\author{A.~\surname{Kim}}
\affiliation{Department of Physics, Sungkyunkwan University, Suwon 440-746, Korea,}
\author{R.~L.~\surname{Kozub}}
\affiliation{Department of Physics, Tennessee Technological University, Cookeville, Tennessee 38505, USA,}
\author{E.~J.~\surname{Lee}}
\affiliation{Department of Physics, Sungkyunkwan University, Suwon 440-746, Korea,}
\author{B.~\surname{Manning}}
\affiliation{Department of Physics and Astronomy, Rutgers University, New Brunswick, New Jersey 08903, USA,}
\author{M.~\surname{Matos}}
\affiliation{Department of Physics and Astronomy, Louisiana State University, Baton Rouge, LA 70803, USA,}
\author{P.~D.~\surname{O'Malley}}
\affiliation{Department of Physics and Astronomy, Rutgers University, New Brunswick, New Jersey 08903, USA,}
\author{S.~D.~\surname{Pain}}
\affiliation{Physics Division, Oak Ridge National Laboratory, Oak Ridge, Tennessee 37831, USA,}
\author{W.~A.~\surname{Peters}}
\affiliation{Oak Ridge Associated Universities, Oak Ridge, Tennessee 37831, USA,}
\author{S.~T.~\surname{Pittman}}
\affiliation{Physics Division, Oak Ridge National Laboratory, Oak Ridge, Tennessee 37831, USA,}
\author{A.~\surname{Ratkiewicz}}
\affiliation{Department of Physics and Astronomy, Rutgers University, New Brunswick, New Jersey 08903, USA,}
\author{M.~S.~\surname{Smith}}
\affiliation{Physics Division, Oak Ridge National Laboratory, Oak Ridge, Tennessee 37831, USA,}
\author{S.~\surname{Strauss}}
\affiliation{Department of Physics and Astronomy, Rutgers University, New Brunswick, New Jersey 08903, USA,}

\date{\today}

\begin{abstract}

The $^{24}$Mg($p$, $\alpha$)$^{21}$Na reaction was measured at the Holifield Radioactive Ion Beam Facility at Oak Ridge National Laboratory in order to better constrain spins and parities of energy levels in $^{21}$Na for the astrophysically important $^{17}$F($\alpha, p$)$^{20}$Ne reaction rate calculation. 31 MeV proton beams from the 25-MV tandem accelerator and enriched $^{24}$Mg solid targets were used. Recoiling $^{4}$He particles from the $^{24}$Mg($p$, $\alpha$)$^{21}$Na reaction were detected by a highly segmented silicon detector array which measured the yields of $^{4}$He particles over a range of angles simultaneously. A new level at 6661 $\pm$ 5 keV was observed in the present work. The extracted angular distributions for the first four levels of $^{21}$Na and Distorted Wave Born Approximation (DWBA) calculations were compared to verify and extract angular momentum transfer.
\end{abstract}

\maketitle

\section{Introduction}

Detecting $\gamma$-rays from the decay of the long-lived radionuclide $^{44}$Ti ($t_{1/2}$=59.1y) provides a direct calibration of the nucleosynthesis in core-collapse supernovae \cite{comptel}. Because of its importance, the abundance of $^{44}$Ti has been the focus of many studies \cite{comptel,ti01,ti02,ti03,ti04,ti05,17f}. As an example, G. Magkotsios \emph{et al.} \cite{17f} investigated the $^{44}$Ti abundance produced from core-collapse supernovae by studying the impact on the $^{44}$Ti abundance evolution of variation of the nuclear reactions, including ($\alpha, \gamma$), ($\alpha, p$), ($p, \gamma$), ($p, \alpha$), ($p, n$), and ($\alpha, n$) in light and intermediate - mass targets. In their sensitivity study, it was found that the variation in the $^{17}$F($\alpha, p$)$^{20}$Ne reaction rate causes a ``primary" impact on the $^{44}$Ti abundance. The $^{17}$F($\alpha, p$)$^{20}$Ne reaction rate, however, has never been measured. Because the reaction rate may be dominated by the properties of energy levels of $^{21}$Na above the $\alpha$-threshold at 6.561 MeV, searching for energy levels of $^{21}$Na and studying their properties may impact our understanding of the abundance evolution of $^{44}$Ti.

The $^{24}$Mg($p, \alpha$)$^{21}$Na reaction was measured at the Holifield Radioactive Ion Beam Facility (HRIBF) at Oak Ridge National Laboratory (ORNL) in order to make a spectroscopic study of the energy levels in the $^{21}$Na for the $^{17}$F($\alpha, p$)$^{20}$Ne reaction rate at stellar temperatures. The $^{24}$Mg($p, \alpha$)$^{21}$Na reaction was reported only once previously, by J. G. Pronko \textit{et al.} \cite{ref04}, where the lower excited states ($E_x <$ 5 MeV) of $^{21}$Na were studied. Spins and parities of three excited states located at 0.332, 1.716, and 2.829 MeV were constrained by analyzing particle-$\gamma$-ray angular correlations. In the present work, energy levels of $^{21}$Na up to $E_x$ $\sim$ 8.5 MeV were observed, which will provide the first spectroscopic results of the $^{24}$Mg($p$, $\alpha$)$^{21}$Na reaction in the energy range of $E_x$ = 5-8.5 MeV. An partial analysis of the data, for the lowest four states, is presented in this contribution.

\section{EXPERIMENT}

\begin{figure}
  \includegraphics[width=10cm]{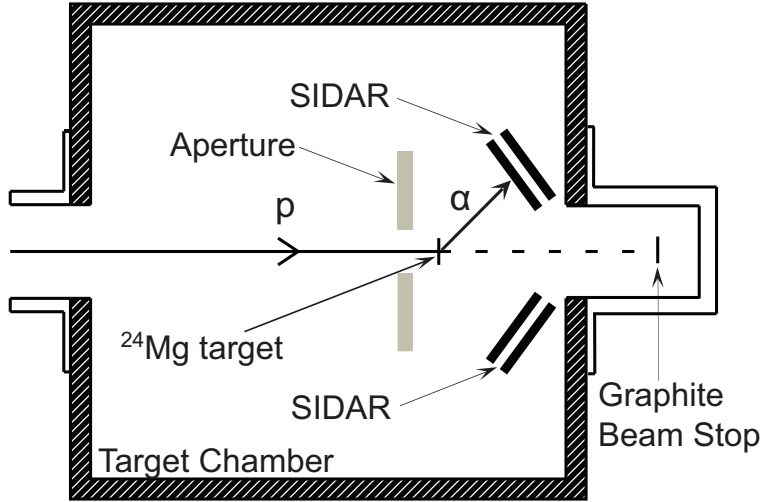}
  \centering
  \caption{A schematic diagram of the experimental setup is shown. A 31 MeV proton beam impinged on an enriched $^{24}$Mg solid target. Recoiling $\alpha$ particles from the $^{24}$Mg(p, $\alpha$)$^{21}$Na reaction were detected at forward angles by the SIDAR. A thick graphite beam stop was used for beam current monitoring. }
  \label{expset}
\end{figure}

A schematic diagram of the experimental setup is shown in Fig. \ref{expset}. A 31 MeV proton beam from the 25-MV tandem accelerator bombarded a pure $^{24}$Mg solid target placed at one of two different target positions at 3.75 in. and 1.75 in. from the front side of the silicon detector array to subtend two separate but overlapping angular ranges. The targets were isotopically enriched (99.9 \%) and had thicknesses of 520-$\mu$g/cm$^2$ and 516-$\mu$g/cm$^2$, respectively. A 0.375 in. thick aluminum plate with an aperture ($\sim$ 0.75 in. diameter) was placed just upstream of the target ladder in order to prevent the fragile silicon detectors from being exposed to the intense proton beam.

Alpha particles from the $^{24}$Mg($p$, $\alpha$)$^{21}$Na reaction were detected by a large area annular silicon detector array (SIDAR) \cite{Dan00}. The SIDAR was composed of four trapezoidally shaped $\Delta$$E-E$ telescopes for particle identification. Each telescope was configured with a thin (100-$\mu$m) $\Delta$$E$ detector backed by a thick (1000-$\mu$m) $E$ detector. With this detector thickness and a beam energy of 31 MeV, even the most energetic $\alpha$ particles from the ($p, \alpha$) reactions (i.e., the $\alpha$ particles associated with the ground state of $^{21}$Na) can be fully stopped in the detector system. The detector wedges were tilted upstream 43$^\circ$ from the perpendicular to the beam axis in order to cover a larger angular range. Since each detector is segmented into 16 azimuthal strips, the reaction cross sections were measured at multiple angles simultaneously, and thus the angular distributions of recoiling $\alpha$ particles were extracted without the need for relative normalizations. The angles covered by the SIDAR were 17$^\circ$ $<$ $\theta$$_{lab}$ $<$ 44$^\circ$ (19$^\circ$ $<$ $\theta$$_{c.m.}$ $<$ 50$^\circ$) and 24$^\circ < \theta_{lab} < 63 ^\circ$ (27$^\circ < \theta_{c.m.} < 70 ^\circ$) for the two target positions. By placing two targets at different positions, a wider range of angles could be covered without changing any other parameters of the experimental setup. Three out of four detector telescopes were with respect to gain for the $^{24}$Mg($p, d$)$^{23}$Mg reaction \cite{minsik} which was measured simultaneously. Since the energies of the $\alpha$ particles from $^{24}$Mg($p$, $\alpha$)$^{21}$Na are about factor of 2 higher than those of the deuterons from the $^{24}$Mg($p, d$)$^{23}$Mg reaction, the fourth telescope was optimized in gain for the ($p$, $\alpha$) channel. The beam current was continuously integrated from a thick graphite beam stop which was placed on the downstream side of the target chamber.

\section{DATA ANALYSIS}

\begin{figure}
  \includegraphics[width=10cm]{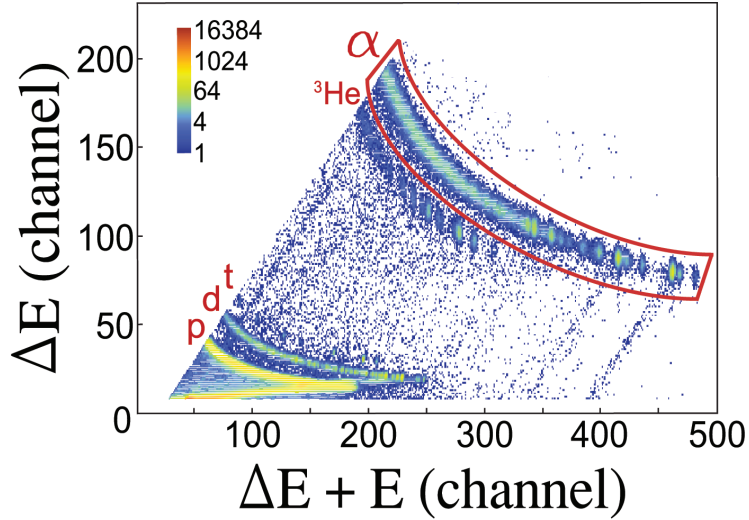}
  \centering
   \caption{(Color online) An example of the energy loss ($\Delta$$E$) versus total energy ($\Delta$$E$ + $E$) plot obtained at $\theta$$_{lab}$ = 27.2$^\circ$ ($\theta$$_{c.m.}$ = 30.4$^\circ$) is shown. Events falling in the red gate were identified as $\alpha$ particles. Other reaction products are also labeled.}
  \label{2dhis}
\end{figure}

To calibrate the energy response of the silicon detectors, an $\alpha$ emitting source composed of $^{237}$Pu (5.157 MeV), $^{241}$Am (5.486 MeV), and $^{244}$Cm (5.805 MeV) was used. By using $\alpha$ peaks of the three energies, the energy offset of each Analog to Digital Converter (ADC) channel could be determined as well. An additional calibrated $^{244}$Cm source was used to measure the solid angle subtended by each strip. The measured solid angles were cross-checked with calculated solid angles using the known detector geometry. The measured solid angles agreed with geometric calculations to within 3 \%.

\begin{figure}

  \includegraphics[width=10cm]{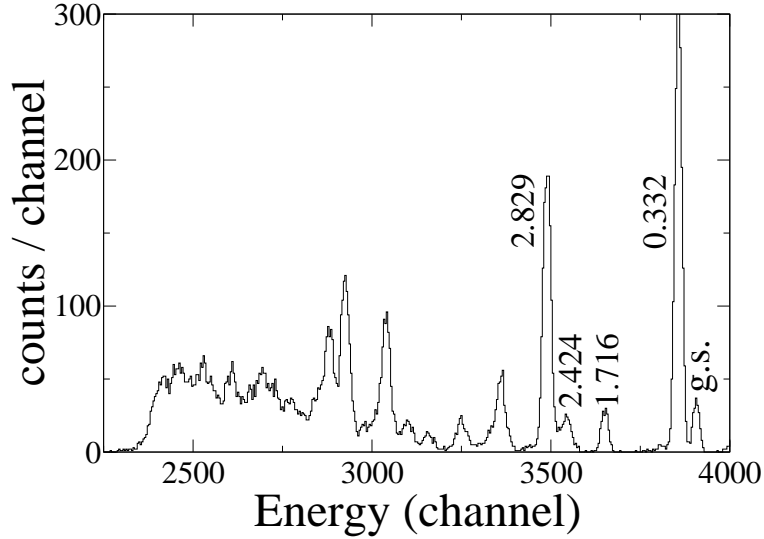}
  \centering
  \caption{The number of counts per channel versus $\alpha$ energy plot at $\theta_{lab}$ = 27.2$^\circ$ is shown. Excitation energies are labeled in MeV. The labeled levels have well-known spin assignments.}
  \label{1dhis}
\end{figure}

Light charged particles were identified by a standard energy loss technique. A typical particle-identification plot is shown in Fig. \ref{2dhis}. The plot was obtained for the strip subtending at $\theta_{lab}$ = 27.2$^\circ$ ($\theta$$_{c.m.}$ = 30.4$^\circ$) in this case. Events falling in the red gate in Fig. 2 were identified as $^{4}$He particles. The $^{4}$He yields were clearly identified as shown in the figure without significant evidence of contamination from the other charged particle groups such as $p$, $d$, $t$, and $^{3}$He. The total energies of the $^{4}$He particles were reconstructed by summing the energies deposited in the $\Delta$$E$ and $E$ detectors. Fig. \ref{1dhis} shows an example of a total energy spectrum of gated $^{4}$He particles from the particle identification plot. Many energy levels of the radioactive $^{21}$Na nucleus are evident in this spectrum.

Because the recorded ADC channels are not always linearly dependent on the actual energies of incident particles, internal energy calibrations are necessary to more precisely determine the energy spectra. The internal calibrations were performed at each angle using two energy levels of $^{21}$Na located at 0.332 MeV and 2.829 MeV excitation, both of which were strongly populated as shown in Fig. \ref{1dhis}. Calibrated $^{4}$He energies were then converted to excitation energies in $^{21}$Na using the well-known detector geometry and reaction kinematics. Five levels, including the ground state, were clearly identified as labeled in Fig. \ref{1dhis}. As shown in the figure, however, many excited states were populated as well at the energies above $E_x >$ 2.829 MeV. Improved internal energy calibrations are required to determine precise excitation energies of the higher-lying states due to the required extrapolation to low $^{4}$He energies.

\begin{figure*}[t]
  \begin{center}
  \includegraphics[width=15cm]{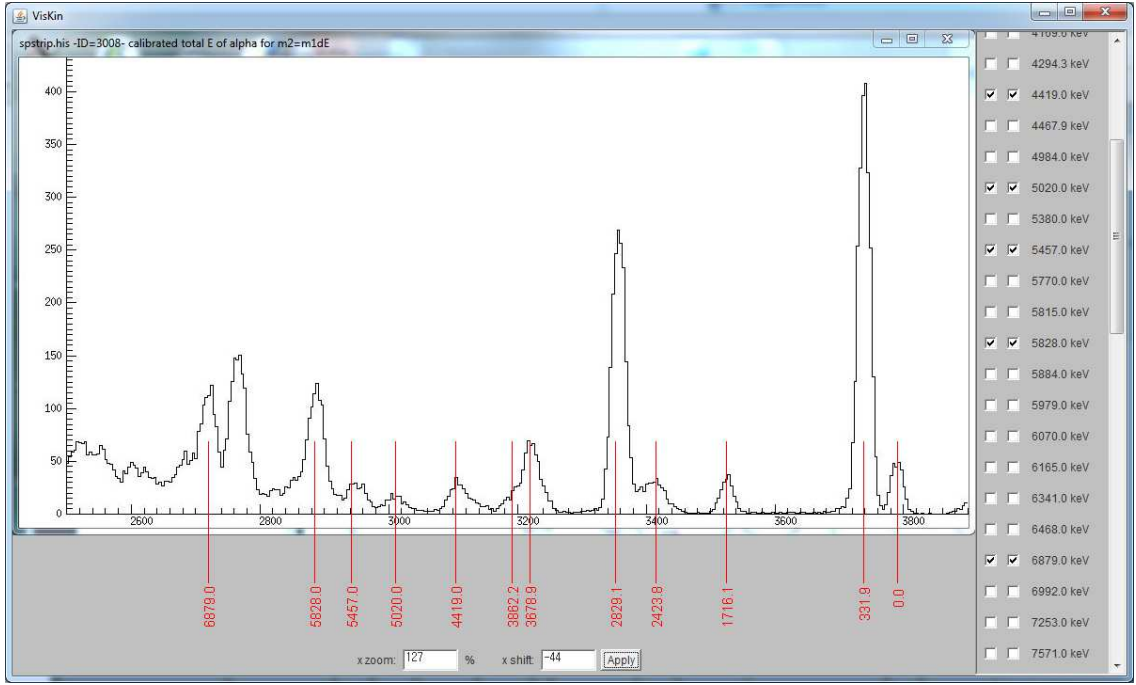}
  \centering
  \caption{(Color online) The result of internal calibrations performed at $\theta_{lab}$ = 29.1$^\circ$ data are shown.}
  \label{viskin}
  \end{center}
\end{figure*}

In order to verify the results of the internal calibrations, a newly developed relativistic kinematics calculation code VisKin (Visualized Kinematics) was used \cite{EJ14}. By using the object-oriented computer programming language Java \cite{java01}, VisKin helps users to interact with the calculated results so that the internal calibrations can be performed relatively easily. Upon execution, VisKin requires users to upload an image of an energy spectrum that will be used for the internal calibration. Information regarding the nuclear reaction such as masses and atomic numbers of particles participating in the reaction, beam energy, and emitting laboratory angle of the ejectile is also required to perform the relativistic kinematics calculations. The program then reads in the excitation energies of the heavy recoil, which are extracted from the Evaluated Nuclear Structure Data File (ENSDF) database provided by the National Nuclear Data Center of the Brookhaven National Laboratory \cite{NNDC}. After calculating the energies of the ejectiles that would be produced by the population of known energy levels of the heavy recoils, the program displays the energies as red lines on top of the energy spectrum image which the user provided as an input. The calculated results (red lines) can be translated and zoomed so that the internal calibrations can be determined easily (Fig. \ref{viskin}).

\begin{figure}
  \includegraphics[width=10cm]{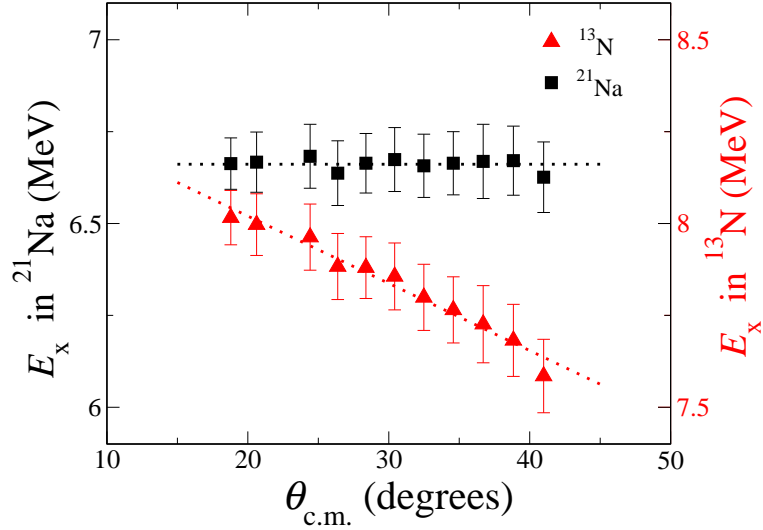}\\
  \centering
  \caption{(Color online) Excitation energies extracted at 11 angles for a peak located near channel number 2770 in Fig. \ref{viskin} are plotted as a function of $\theta_{c.m.}$. The black squares (red triangles) represent the calculated excitation energies of $^{21}$Na ($^{13}$N) assuming the peak arose from the $^{24}$Mg($p$, $\alpha$)$^{21}$Na ($^{16}$O($p$, $\alpha$)$^{13}$N) reaction.}
  \label{6661}
\end{figure}

The result of internal calibrations performed for the $^{24}$Mg($p$, $\alpha$)$^{21}$Na reaction study is shown in Fig. \ref{viskin}. The experimental data represent counts per channel obtained at $\theta_{lab}$ = 29.1$^\circ$ as a function of the total $\alpha$ energy. As shown in the figure, lower-lying levels of $^{21}$Na up to $E_{x}$ $\sim$ 6.9 MeV were well reproduced by known energy levels from references \cite{ref01,ref02,ref03,ref04,ref05} except the level located near channel number 2770. The excitation energy in $^{21}$Na for this level was calculated to be 6661 $\pm$ 5 keV, which does not correspond to any known $^{21}$Na levels. The excitation energy uncertainty was calculated by using Eq. (1) of Ref. \cite{chae10}. Excitation energies extracted at 11 angles are plotted in Fig. \ref{6661} as a function of $\theta_{c.m}$. As shown in the figure, the excitation energies are rather consistent indicating that the peak did not arise from any contamination in the target.

The differential cross sections of the identified levels in the center of mass system at each angle were calculated as
\begin{equation}
\label{cross section}
  \left( \frac{d\sigma}{d\Omega} \right)_{j,\theta}=\frac{Y_{j,\theta}}{IN\Delta\Omega_{\theta}},
\end{equation}
where $Y_{j,\theta}$ is the yield of $\alpha$ particles from the $^{24}$Mg($p$, $\alpha$)$^{21}$Na reaction for an energy level $j$ emitted at an angle of $\theta$ in the center of mass system, $I$ is the number of beam particles incident on the target, $N$ is the number of target atoms per unit area, and $\Delta\Omega_{\theta}$ is the solid angle covered by a strip of SIDAR in the center of mass system.

Angular distributions of four levels located at energies of 0.0, 0.332, 1.716, and 2.424 MeV were extracted from the experimental data. Since these four levels have well-known spin and parity assignments \cite{ref01, ref02, ref03, ref04, ref05}, comparing experimental angular distributions of the levels with Distorted Wave Born Approximation (DWBA) calculations considering proper $l$ transfer will verify the validity of the optical model parameters used in the calculations. DWBA calculations were performed using the zero range computer code DWUCK4 \cite{dwuck}. The optical parameters were adopted from previous work \cite{flem71} and slightly modified to better fit experimental data. If extracting spectroscopic factors is the main purpose of the study, fine-tuning optical model parameters obtained from previous scattering experiments would be inappropriate. In the present work, however, the goal is to find the proper $l$ transfer that is consistent the maxima and minima observed in the experimental angular distribution. Table. \ref{opt} summarizes the optical model parameters used for the calculations.

\begin{table*}
\caption{\label{opt} The Optical model parameters used for DWUCK4 code in this work are shown. The definitions of these parameters follow the conventions in the Ref. \cite{dwuck}.}
\begin{ruledtabular}
\begin{tabular}{cccccccc}

  Particle & VR (MeV) & r$_{0R}$ (fm) & AR (fm) & VI (MeV) & r$_{0I}$ (fm) & AI (fm) & r$_{0C}$ (fm) \\
  \hline
  p & 44.53 & 1.141 & 0.15 & 17.51 & 1.26 & 0.64 & 1.3 \\
  $^{4}$He & 90.1 & 1.2 & 0.87 & 13.8 & 1.8 & 0.99 & 1.3 \\

\end{tabular}
\end{ruledtabular}
\end{table*}

\begin{figure}

  \includegraphics[width=10cm]{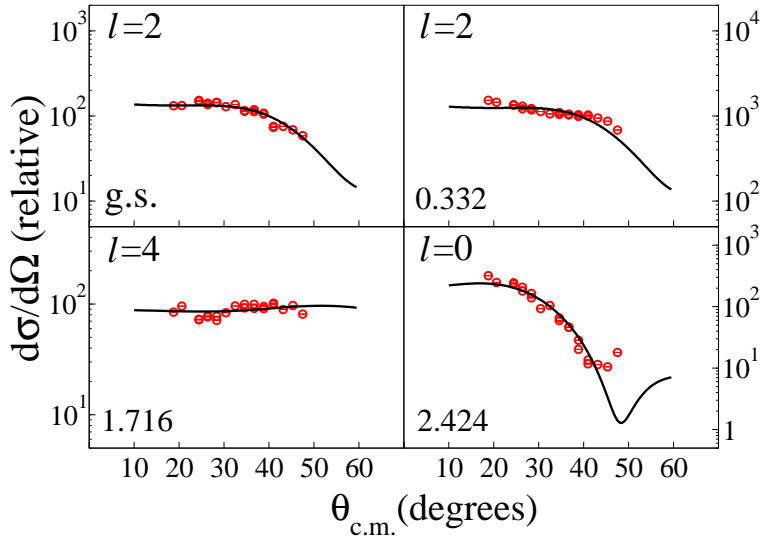}\\
  \centering
  \caption{(Color online) The angular distributions of $\alpha$ particles from the $^{24}$Mg($p$, $\alpha$)$^{21}$Na reaction (circles) and best-fitting DWBA calculations (black solid lines) for four levels are shown. The statistical uncertainties of differential cross sections are smaller than the size of the circles. All excitation energies, labeled on the bottom-left of each plot, are in MeV.}
  \label{angular}
\end{figure}

Fig. \ref{angular} shows the angular distributions of $^{4}$He particles for the first four levels including the ground state. The circles represent experimental data obtained from this work, and black solid lines represent DWBA calculations that best fit the experimental angular distributions. Statistical uncertainties are marked as error bars. In most cases, however, the uncertainties are smaller than the size of data points. Triton transfer was assumed for the calculations. Transferred $l$ values are also shown in each plot. The number of beam particles incident on the target ($I$) and the number of target atoms per unit area ($N$) in Eq. (\ref{cross section}) are not included in the calculations of the experimentally obtained differential cross sections. This does not affect the shape of the angular distributions, since the cross sections were measured at all angles simultaneously. As shown in Fig. \ref{angular}, the DWBA calculations well reproduce the extracted angular distributions for the four well-known energy levels.

\section{CONCLUSION AND FUTURE PLAN}

A spectroscopic study of the $^{24}$Mg($p$, $\alpha$)$^{21}$Na triton transfer reaction was made by using 31 MeV proton beams at the HRIBF of the ORNL to constrain spins of energy levels in $^{21}$Na for the astrophysically-important $^{17}$F($\alpha$, $p$)$^{20}$Ne reaction rate. The energies and angular distributions of recoiling $^{4}$He particles from the reaction were measured using a highly segmented silicon detector array. By comparing experimental differential cross sections of strongly populated $^{21}$Na energy levels with theoretical DWBA calculations, we could verify the following spin and parity assignments: ground state - $\frac{3}{2}^{+}$, $E_{x}$=0.332 MeV - $\frac{5}{2}^{+}$, $E_{x}$=1.716 MeV - $\frac{7}{2}^{+}$, $E_{x}$=2.424 MeV - $\frac{1}{2}^{+}$, which supports the validity of the optical model parameters used for the calculations.

Many energy levels in $^{21}$Na up to $\sim$ 6.9 MeV were observed, some of which do not have definite spin assignments and others have not been observed previously. To determine precise excitation energies and to constrain spins and parities of the higher-lying states, improved internal calibrations are required. Analysis of the experimental data obtained at the second target position will extend the angular coverage up to $\theta$$_{c.m.}$ $\sim$  70$^\circ$, which will be useful to constrain the spin assignments. The $^{17}$F($\alpha$, $p$)$^{20}$Ne reaction rate will be updated at stellar temperatures as well.

\begin{acknowledgments}
This work was supported by a National Research Foundation of Korea (NRF) grant funded by the Korea government Ministry of Education, Science, and Technology (MEST) No. NRF-2014S1A2A2028636. This research was supported in part by the National Nuclear Security Administration under the Stewardship Science Academic Alliances program through U.S.DOE Cooperative Agreement No. DE-FG52-08NA28552 with Rutgers University and Oak Ridge Associated Universities. This work was also supported in part by the Office of Nuclear Physics, Office of Science of the U.S.DOE under Contracts No. DE-FG02-96ER40955 with Tennessee Technological University, No. DE-FG02-96ER40983 with the University of Tennessee, and DE-AC-05-00OR22725 at Oak Ridge National Laboratory; and the National Science Foundation.
\end{acknowledgments}

\end{document}